\documentclass{article}

\usepackage{PRIMEarxiv}

\usepackage[utf8]{inputenc} 
\usepackage[T1]{fontenc}    
\usepackage{hyperref}       
\usepackage{url}            
\usepackage{booktabs}       
\usepackage{amsfonts}       
\usepackage{nicefrac}       
\usepackage{microtype}      
\usepackage{lipsum}
\usepackage{fancyhdr}       
\usepackage{graphicx}       
\graphicspath{{media/}}     
\usepackage{booktabs}
\usepackage{multirow}
\usepackage{wrapfig}
\usepackage{floatrow}

\newfloatcommand{figurebox}{figure}[\nocapbeside][\dimexpr(\textwidth-\columnsep)/2\relax]
\newfloatcommand{tablebox}{table}[\nocapbeside][\dimexpr(\textwidth-\columnsep)/2\relax]

\usepackage{soul}
\usepackage{color, xcolor}
\usepackage{upgreek}
\usepackage{graphicx}
\usepackage{amsmath,amsthm,amssymb,amsfonts,bm}
\usepackage{mleftright,mathtools}
\usepackage{svg}
\usepackage{stfloats}
\usepackage{indentfirst}
\usepackage{multirow}
\usepackage{multicol}
\usepackage{makecell}
\usepackage{setspace}

\usepackage{amssymb}
\usepackage{graphicx}
\usepackage{array}
\newcolumntype{C}[1]{>{\centering\arraybackslash}p{#1}}
\newcolumntype{L}[1]{>{\arraybackslash}p{#1}}
\newcolumntype{R}[1]{>{\raggedleft\arraybackslash}p{#1}}

\usepackage{algorithm}
\usepackage{algorithmic}
\usepackage{amsmath}

\usepackage{tabularx} 
\usepackage{booktabs} %

\usepackage{scrextend}
\usepackage{microtype}   

\usepackage[accsupp]{axessibility}
\title{BEV-GS: Feed-forward Gaussian Splatting in Bird's-Eye-View for Road Reconstruction
}

\usepackage{authblk}

\author{
Wenhua Wu$^{1}$ \qquad  Tong Zhao$^{2*}$\qquad Chensheng Peng$^{3}$ \qquad Lei Yang$^{2}$ \qquad Yintao Wei$^{2}$\\
Zhe Liu$^{1}$ \qquad Hesheng Wang$^{1}\thanks{Corresponding authors.}$ \vspace{0.5em}\\
$^{1}$Shanghai Jiao Tong University \qquad $^{2}$Tsinghua University\qquad
$^{3}$University of California, Berkeley
}
\begin{document}
\maketitle
\vspace{-0.3in}
\begin{figure*}[h] 
\center{
\includegraphics[width=\linewidth]{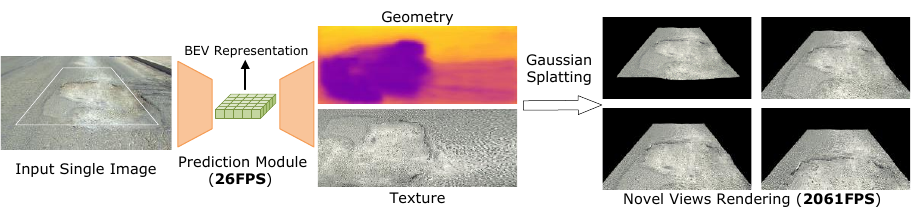}}
\caption{ 
\textbf{BEV-GS} is capable of real-time road surface reconstruction from a single image. Unlike existing per-pixel feed-forward Gaussian prediction methods, it predicts geometry and texture parameters of mesh grid Gaussians based on BEV features. BEV-GS achieves remarkable road surface reconstruction performance and can rapidly render novel views of the road. 
        The images on the right side simulate the captured road as the car moves forward.
        }
    \label{fig:teaser}
\end{figure*}

\begin{abstract}

Road surface is the sole contact medium for wheels or robot feet. Reconstructing road surface is crucial for unmanned vehicles and mobile robots. Recent studies on Neural Radiance Fields (NeRF) and Gaussian Splatting (GS) have achieved remarkable results in scene reconstruction.
However, they typically rely on multi-view image inputs and require prolonged optimization times. 
In this paper, we propose BEV-GS, a real-time single-frame road surface reconstruction method based on feed-forward Gaussian splatting. BEV-GS consists of a prediction module and a rendering module. The prediction module introduces separate geometry and texture networks following Bird's-Eye-View paradigm. Geometric and texture parameters are directly estimated from a single frame, avoiding per-scene optimization. In the rendering module, we utilize grid Gaussian for road surface representation and novel view synthesis, which better aligns with road surface characteristics. Our method achieves state-of-the-art performance on the real-world dataset RSRD. The road elevation error reduces to 1.73 cm, and the PSNR of novel view synthesis reaches 28.36 dB. The prediction and rendering FPS is 26, and 2061, respectively, enabling high-accuracy and real-time applications. 
The code will be available at: 
\href{https://github.com/cat-wwh/BEV-GS}{\texttt{https://github.com/cat-wwh/BEV-GS}} \\

\end{abstract}    

\section{Introduction}
\label{sec:intro}

Road surface plays a vital role in unmanned vehicles and robots, as it is the only contact media between wheels and the physical world.
Geometry and texture information are the most significant road properties. Road geometry or elevation determines ride comfort and motion smoothness. Road texture and color reflect material, friction level, and unevenness, which are crucial clues for understanding road scene conditions. 

Road surface reconstruction (RSR), aiming at recovering the properties above, significantly improves the driving safety and comfort of autonomous driving vehicles~\cite{THEUNISSEN2021206,10329453}. Online road preview with camera sensors is a promising way that gains attention. Reconstruction methods based on monocular and multi-view images in perspective view (PV) have been extensively explored \cite{9025600,mei2023rome}. They recover road point clouds from dense depth estimation associated with direct color projection. However, unstructured road representation inhibits downstream applications or fusing multi-modal data. RoadBEV \cite{roadbev} first proposes to recover road profiles in Bird's Eye View (BEV) with both monocular and stereo settings. Nevertheless, they reconstruct only road geometry without considering color.

Recent studies on Neural Radiance Fields (NeRF) \cite{mildenhall2021nerf} and Gaussian Splatting (GS) \cite{kerbl20233d} have achieved remarkable results in large-scale scene reconstruction, particularly in color rendering. However, these methods typically rely on multi-view image inputs and require long optimization time \cite{ost2021neural, turki2023suds, zhou2024drivinggaussian}. EMIE-MAP \cite{wu2024emie} and RoGS \cite{feng2024rogs} introduce implicit representations and GS into road surface reconstruction. They are offline and scene-specific, requiring iterative optimization on input image sequences. Additionally, they rely on structure from motion (SfM) or vehicle trajectory information for initializing road surface geometry. 

The vanilla NeRF and GS methods memorize a certain scene from massive images. Despite the strong fitting capabilities, they are not generalizable to new scenarios since universal prior is not learned. Feed-forward GS-based methods have been developed for single frame scene reconstruction \cite{szymanowicz2024splatter_image, szymanowicz2024flash3d}. Specifically, they use an image-to-image network to predict per-pixel GS parameters and then render novel view images. With sufficient prior information learned during training, this paradigm is scene-independent and generalizable to unseen scenarios. However, existing methods rely on pre-trained depth estimation models. Per-pixel prediction is not suitable for scenes with large near-far distance spans like open streets. The perspective effect causes near points to be dense and far points to be sparse.

To resolve the identified issues above, we propose BEV-GS, a real-time road surface reconstruction method combining BEV paradigm and GS, without the need for multi-view inputs and iterative optimization. In contrast to existing pixel-wise feed-forward GS methods, we introduce a novel Gaussian prediction network based on BEV feature grids, where scene geometry and texture are decoupled. For road geometry, image features in PV are transformed into 3D road voxels of interest in BEV, which facilitates surface occupancy prediction for every grid. An independent backbone is introduced to extract texture features. The recovered road elevation contributes to more accurate texture feature query, based on which Gaussian-related color parameters are decoded for grids. Building upon the predictions, we construct a road surface representation based on grid Gaussian. Gaussians conditioned on the accurate surface profile naturally fit road scene, avoiding complex geometric regularization. Predictive and realistic novel view synthesis is achieved in real-time. Benefiting from the BEV representation and decoupled texture and geometry, its superiority is sufficiently verified by experiments on real-world dataset.

Our main contributions are summarized as follows:
\begin{itemize}
    \item We propose BEV-GS, a feed-forward Gaussian Splatting method based on BEV representation with application to road surface reconstruction. It addresses the issue of uneven information distribution in perspective view.
    \item We introduce decoupled geometry and texture prediction. It enables independent optimization, while the geometry branch promotes more accurate texture feature query. 
    \item We achieved state-of-the-art performance on real-world road surface reconstruction dataset, showcasing its capability in real-time road surface reconstruction and novel view synthesis.
\end{itemize}

\section{Related Works}
\label{sec:related}

\noindent {\bf Road surface reconstruction.} RSR is also an essential topic in pavement engineering and remote sensing \cite{600079, GUO2015165}. RSR for unmanned vehicles and robots concentrates more on surface profile, texture, and semantics recovery with affordable LiDAR \cite{ni2020road} or camera \cite{60900001, 7797253, WOS:001025464800002}, aiming at improving safety and smoothness. LiDAR directly scans unstructured road profile, while lacks texture information~\cite{WOS:000930950600001,zhao2024road}. RoadBEV reconstructs road geometry in BEV with both monocular and stereo cameras \cite{roadbev}, achieving errors at 1.83 cm and 0.50 cm, respectively. RoME first utilizes road mesh representation and reconstructs large-scale road with NeRF from sequence images, offering both road structure and color for 4D labeling \cite{mei2023rome}. Further, EMIE-MAP predicts elevation and color with independent MLPs, achieving novel view synthesis with direct projection \cite{wu2024emie}. Nevertheless, road geometry suffers from poor accuracy as road profiles are unsupervised or supervised by sparse labels.

\begin{figure*}[t] 
\center
\includegraphics[width=1.0\textwidth]{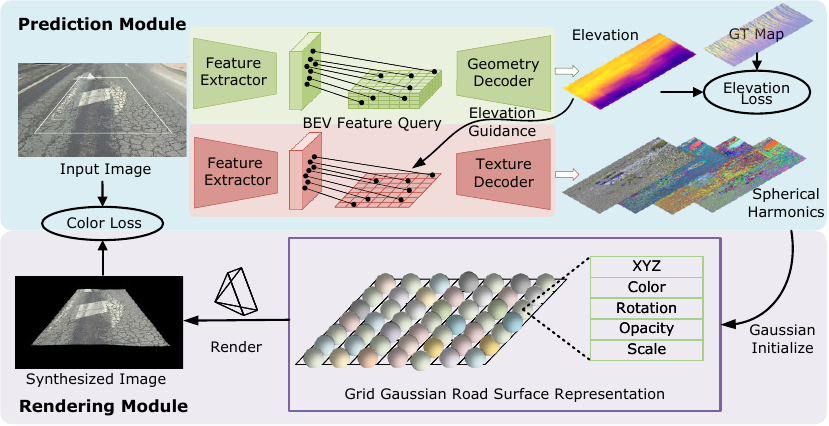}
\caption{Overview of BEV-GS. It consists of a prediction module and a rendering module. In the prediction module, two independent geometry and texture branches based on BEV features predict road elevation and spherical harmonic parameters.
In the rendering module, we represent road surface  with grid Gaussians, which are initialized by the predicted properties. Elevation and color losses separately supervise the feed-forward predictions.
}
\label{framework}
\end{figure*}

\noindent {\bf Reconstruction with Gaussian Splatting.} GS with explicit scene representation is becoming popular for its impressive efficiency \cite{chen2024survey,wu20244d}. Its improved forms have been applied to various scenarios like street view~\cite{yu2024sgd}, dynamic scene~\cite{Zhou_2024_CVPR} and SLAM~\cite{Matsuki_2024_CVPR}. The Gaussians are generally initialized on points derived from SfM or depth estimation. Despite the density adjustment, geometry still suffers from poor performance like floaters. Most recent works focus on geometric regularization for smooth and consistent surface \cite{fan2024trim,zhang2024fregs,chen2023neusg}. GS2Mesh~\cite{wolf2024gsmesh} extracts smooth mesh from noisy points with the aid of stereo images. Trim GS~\cite{fan2024trim} proposes a trimming strategy that removes redundant or inaccurate Gaussians. Despite the efforts, Gaussian centers are still presented in an unstructured style. Adaptive regularization strategies for universal scenes are challenging.

\noindent {\bf Feed-forward conditioned rendering.} To tackle the challenge of scene extrapolation and generalization in vanilla NeRF and GS methods, feature-based rendering with scene encoding is introduced. Works like \cite{muller2022instant, kulhanek2023tetra, sun2022direct} treat grid-based features as parameters and optimize them by back-propagation, which still have the identified problems above. The pixelNeRF \cite{yu2021pixelnerf} first propose a learning-based feed-forward framework for predicting NeRFs from single or several images. GSN \cite{GSN2021} and RNRMap \cite{RNRMap2023} map image features extracted by CNN to 2D floor-plan grids of local latent codes. Given a novel view, camera rays sample latent features by vertical projection and interpolation, based on which a feature map is rendered. 
The Splatter Image \cite{szymanowicz2024splatter_image} maps input image to another image that holds the parameters of one coloured 3D Gaussian per pixel, achieving super fast single-view reconstruction.

\section{Methods}
\subsection{Framework Overview} 
\label{overview}

We propose BEV-GS, a single-frame and feed-forward GS model in BEV with application to road reconstruction. First, following RoadBEV \cite{roadbev}, we predefine a horizontal region of interest in front of the vehicle with size $1.9{\rm m} \times 5.0{\rm m}$. The rectangular area is then discretized into grids, which sufficiently cover the road that ego-vehicle will pass through. Every grid in the mesh representation acts as the minimal unit for road geometry and texture reconstruction.

As illustrated in Figure~\ref{framework}, BEV-GS comprises a prediction module and a rendering module. The former follows the paradigm of BEV perception for representing road surface. Aiming at scene generalization with few shots, we adopt the feed-forward scheme for recovering both road geometry and texture. Two independent backbones are introduced to encode road image and transform into grid or voxel features. The geometry decoder first estimates one-dimensional elevation values of the grids. The recovered 3D road profile facilitates view transformation and feature query of the texture branch. The texture decoder then predicts color-related parameters for every grid in BEV. 
 
In the rendering module, we leverage the grid Gaussians for reconstructing road texture and synthesizing RGB images. With parameters predicted by the texture decoder, 3D Gaussians conditioned on the recovered 3D meshes are splattered across the image plane with differentiable rendering. The texture branch can therefore be end-to-end optimized with the loss between synthesized and actual images. The feed-forward scheme enables predictive novel view synthesis with single-frame input. 

Benefiting from the unified and structured BEV representation, the uneven distribution of information caused by the perspective effect is mitigated. The proposed framework facilitates separate training since the two branches interact only by geometry initialization. The decoupled design enables each branch to focus on its respective domain. 

\subsection{Road Geometry Reconstruction} \label{geometry}
Given a single-frame input $\boldsymbol{I} \in \mathbb{R}^{3\times H\times W}$, the simplified EfficientNet~\cite{efficientnet} with pre-trained weights is adopted as the encoder $E_{g}$ to extract road geometric features $\boldsymbol{F}^{g}_{\rm pv} \in \mathbb{R}^{C\times \frac{H}{4}\times \frac{W}{4}}$. Instead of using transformer-based cross-attention to enlarge the receptive field during BEV query~\cite{park2024heightlane,li2022bevformer}, we adopt direct 3D-2D projection for view transformation since dense road voxels are already set for surface occupancy prediction. The projected reference points aggregate features of nearby pixels with bi-linear weighting. A 4D voxel feature $\boldsymbol{F}_{\rm vox}\in \mathbb{R}^{C\times N^g_x\times N^g_y\times N^g_z}$ is derived, where $N^g_x$, $N^g_y$, and $N^g_z$ denotes the number of anchor voxels in the lateral, longitudinal and vertical directions, respectively.

RoadBEV \cite{roadbev} reduces the height dimension of voxel feature by direct channel concatenation. Instead, to reduce computation while preserving key features, we adopt weighted sum for fusing anchor features of grids. Stacked 3D convolution layers followed by a \emph{Softmax} operation along the height dimension are introduced to evaluate the significance of every elevation anchor. The geometry BEV feature $\boldsymbol{F}^{g}_{\rm bev}\in \mathbb{R}^{C\times N^g_x\times N^g_y}$ is derived as:
\begin{equation}
    \boldsymbol{F}^{g}_{\rm bev}(g)=\sum_{z=1}^{N^g_z}{\boldsymbol{F}_{\rm vox}(g, z) \cdot {\rm Softmax}(\boldsymbol{F}^{\prime}_{\rm vox}(g, \cdot))},
  \label{eq:ele_pred}
\end{equation}

\noindent where $g$ indicates a horizontal grid at $(x, y)$, $\boldsymbol{F}^{\prime}_{\rm vox}\in \mathbb{R}^{N^g_x\times N^g_y\times N^g_z}$ is the single-channel voxel feature exported from the 3D convolution layers.

A geometry decoder $D_g$ composed of basic blocks in EfficientNet~\cite{efficientnet} further estimates elevation based on $\boldsymbol{F}^{g}_{\rm bev}$. Since the elevation reference plane is fixed below camera, actual road surface may have holistically deviation due to vehicle pitch vibration. Instead of direct regression, we separately estimate a global reference plane $\hat{H}_{\rm ref}$ and an offset map $\boldsymbol{\hat{H}}_{\rm off}$, which are all in the range of $[ \frac{h_{\rm min}}{2}, \frac{h_{\rm max}}{2}]$, where $h_{\rm min}$ and $h_{\rm max}$ are the lower and upper elevation bounds, respectively. The final elevation map is their combination:

\begin{equation}
    \boldsymbol{\hat{H}}=\hat{H}_{\rm ref} + \boldsymbol{\hat{H}}_{\rm off},
      \label{eq:ele_pred}
\end{equation}

The reference $\hat{H}_{\rm ref}$ is derived by cascaded average pooling, MLP layers, and \emph{Tanh} activation operations on the feature map. Since direct regression is ineffective for model supervision \cite{roadbev}, the offset value for every grid is obtained by classification on offset bins and regression with weighted sum:

\begin{equation}
    \boldsymbol{\hat{H}}_{\rm off}(g)=\sum_{b=1}^{N_b}{e_b\cdot {\rm Softmax}(\boldsymbol{F}^{e}_{\rm bev}(\cdot, g))},
  \label{eq:ele_offset_pred}
\end{equation}

\noindent where $\boldsymbol{F}^{e}_{\rm bev}\in \mathbb{R}^{N_b\times N^g_x\times N^g_y}$ is the final feature map for elevation offset classification, $e_b$ is the elevation value of corresponding bin. $N_b$ is the number of offset bins. 

Compared with existing GS methods that generally utilize COLMAP \cite{schoenberger2016sfm} for point cloud initialization, the proposed geometry branch offers structured points that are naturally suitable for describing road surface. The accurate scene geometry guarantees reliable feature query and texture decoding for predicting Gaussian-related properties.

\begin{figure*}[t] 
\center
\includegraphics[width=1.0\textwidth]{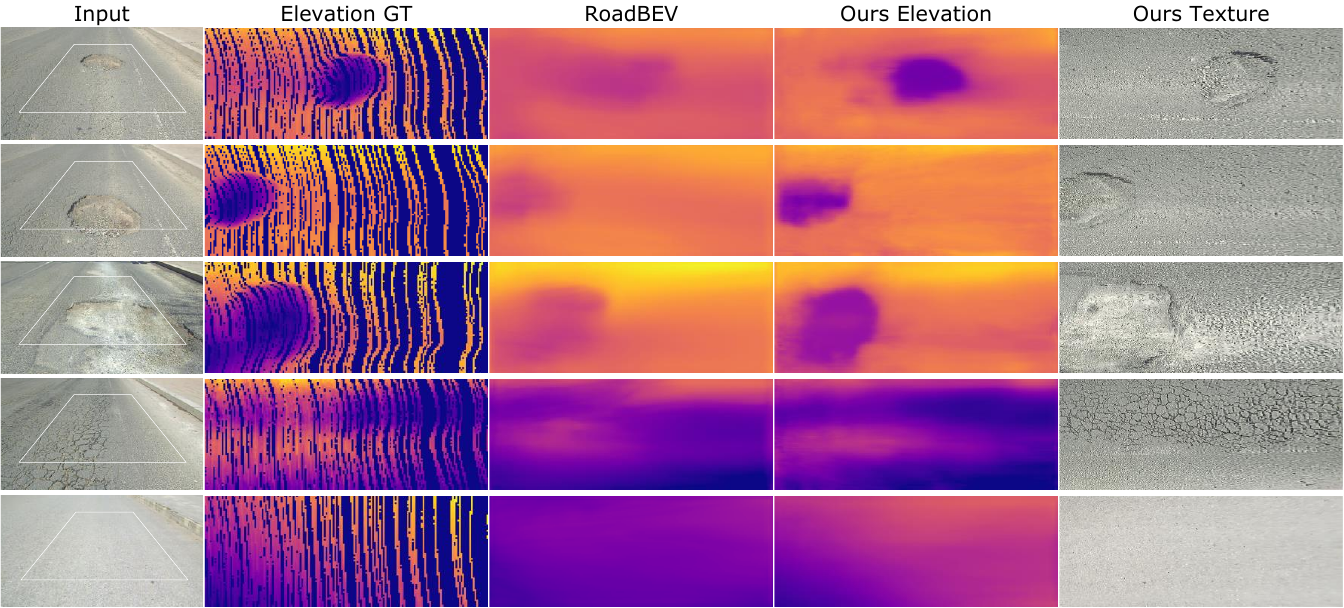}
\caption{Visualization of road reconstruction results under typical scenarios. Compared with RoadBEV \cite{roadbev}, our method not only reconstructs more accurate road geometry but also the detailed road texture.
}
\label{fig-ele}
\end{figure*}

\begin{figure*}[t] 
\center
\includegraphics[width=1.0\textwidth]{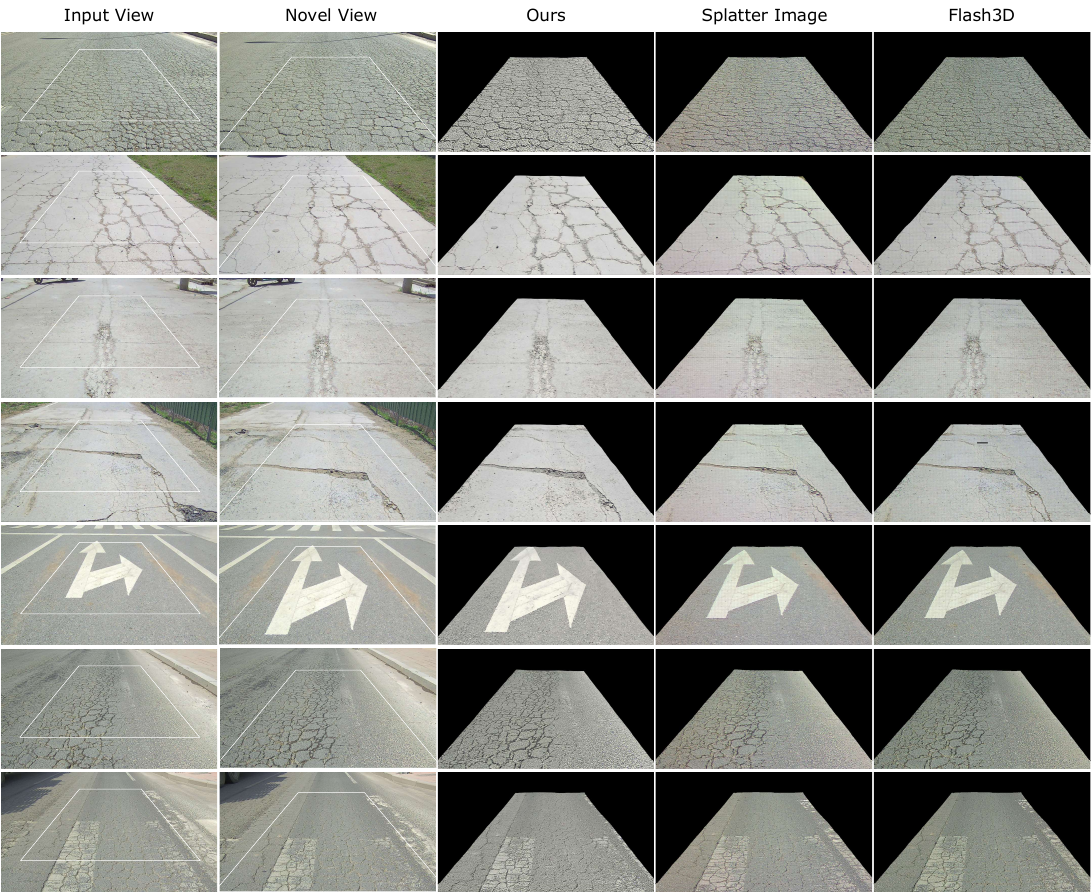}
\caption{Visualization of novel view synthesis, which is the next frame as the vehicle moves forward. For comparison purposes, the render results of Splatter Image \cite{szymanowicz2024splatter_image} and Flash3D \cite{szymanowicz2024flash3d} are restricted to the same area as ours.
}
\label{fig-rgb}
\end{figure*}

\subsection{Road Texture Representations}
The architecture of the texture prediction branch is generally similar to the geometry branch. It also implements a BEV pipeline that predicts texture information for each Gaussian sphere conditioned on the grid.
Unlike in \cite{szymanowicz2024flash3d}, where a shared backbone is used for both geometry and texture prediction, we use a separate texture encoder $E_t$ composed of simplified MobileNetV3~\cite{mobilenetv3}. The texture feature map $\boldsymbol{F}^{t}_{\rm pv} \in \mathbb{R}^{C\times \frac{H}{4}\times \frac{W}{4}}$ with $\frac{1}{4}$ resolution is obtained:

\begin{equation}
    \boldsymbol{F}^{t}_{\rm pv} = E_t(\boldsymbol{I}).
\end{equation}

This decoupled design proves to be more effective as structure and texture recovery require distinct domain features, while shared features suffer from obscure representations. Similarly, we utilize 3D to 2D projection to query texture BEV features $\boldsymbol{F}^{t}_{\rm bev}\in \mathbb{R}^{C\times N^t_x\times N^t_y}$. For more accurate feature queries, we leverage the elevation predicted by the geometry branch for projection guidance without the need for lifting anchors. Subsequently, a texture decoder $D_{t}$ followed by an up-sampling layer is employed to predict dense spherical harmonics coefficients $\boldsymbol{C} \in \mathbb{R}^{ 3(L+1)^2\times 2N^{t}_x\times 2N^{t}_{y}}$ of grids,
where $L$ is the order of the spherical harmonics. The up-sampling aims to obtain more fine-grained and high-resolution road texture information.

\subsection{Feed-forward Gaussian Splatting}

\noindent \textbf{Grid Gaussian road surface representation}
Building upon the BEV-based reconstruction of road geometry and texture, we establish a road surface representation with mesh grid Gaussian. Existing schemes typically maintain non-uniformly distributed 3D/2D Gaussian spheres, necessitating optimization of Gaussian positions and densities. Considering the physical characteristics of road surface, we arrange 3D Gaussian spheres on the uniformly partitioned road grid cells. The uniform distribution of the fixed Gaussians ensures coverage over all road surface areas without the need for density optimization. Each sphere comprises parameters of geometric center, scale, rotation, opacity, and color. The geometric centers are initialized by the horizontal coordinates $(x, y)$ of grid centers and the estimated elevation $\boldsymbol{\hat{H}}(x, y)$. Colors are initialized by predicted spherical harmonic coefficients $\boldsymbol{C}(x, y) \in \mathbb{R}^{3(L+1)^2}$. Scales of Gaussians are all set to $\boldsymbol{S} \in \mathbb{R}^3$. Rotations $\boldsymbol{R} \in \mathbb{R}^{3\times 3}$ are treated as the identity matrix, and opacities are initialized as scalar $\sigma$, making each Gaussian sphere isotropic. The covariance of 3D Gaussians can be expressed as:

\begin{equation}
   \boldsymbol{\Sigma} = \boldsymbol{R}\boldsymbol{S}\boldsymbol{S}^T\boldsymbol{R}^T.
\end{equation}

\begin{table}[t]
  \centering
    \begin{tabular}{c|c||c|c|c}
    \hline
    \multicolumn{2}{c||}{Methods} & \makecell[c]{AAE  (cm) $\downarrow$} & \makecell[c]{RMSE  (cm)$\downarrow$} & \makecell[c]{\textgreater 5mm  (\%)$\downarrow$}\\
    \hline
      \multirow{7}{*}{\rotatebox{90}{PV}} & iDisc \cite{piccinelli2023idisc} & 2.70 & 2.89 & 86.3  \\
      & LapDepth \cite{9316778} &  2.55 & 2.79 & 85.2 \\
      & PixelFormer \cite{Agarwal_2023_WACV} & 2.47 & 2.66 & 84.9  \\
      & AdaBins \cite{9578024} & 2.36 & 2.55 & 84.0  \\
      & ZoeDepth \cite{ZoeDepth} & 2.41 & 2.59 & 82.5  \\
      & {DepthAnything V2 \cite{depth_anything_v2}} & 2.30 & 2.48 & 83.2  \\
      \hline
      \multirow{2}{*}{\rotatebox{90}{BEV}} & RoadBEV \cite{roadbev} & 1.93 & 2.16 & 83.4 \\
      & BEV-GS (Ours) & \textbf{1.73} & \textbf{1.94} & \textbf{80.2} \\
    \hline
    \end{tabular}%
    \caption{Comparison of road geometry reconstruction. Our method shows state-of-the-art performance. \textbf{Bold}: best. `PV' denotes Perspective View.}
    \label{tab-ele}%
\end{table}%

\noindent \textbf{Gaussian Splatting}
Given the camera pose $\boldsymbol{P}$, project the grid Gaussians onto 2D image plane:

\begin{equation}
   \boldsymbol{\Sigma}^\prime = \boldsymbol{J}\boldsymbol{P}\boldsymbol{\Sigma}\boldsymbol{P}^{T}\boldsymbol{J}^{T},
\end{equation}

\noindent where $\boldsymbol{J}$ is the affine-approximated Jacobian matrix of the projection function. The Gaussian sphere distribution is sorted in depth order for rendering color of each pixel $p$:
\begin{equation}
   \hat{\boldsymbol{I}}(p) = \sum_{i \in N}c_i \alpha^{\prime}_i \prod_{j=1}^{i-1}(1-\alpha^{\prime}_j),
\end{equation}

\noindent where $c_i$ represents the color of the $i$-th Gaussian obtained from spherical harmonic coefficients, and $\alpha^{\prime}_i$ represents the density computed from opacity and covariance.

\subsection{Loss Functions} \label{loss}

As emphasized above, the two branches can be trained separately for better performance. For the geometry branch, we use smoothed L1 loss to supervise elevation reconstruction. The elevation loss of a sample is the total loss of grids with valid label: 
\begin{equation}
    \mathcal{L}_{\rm ele}=\sum_{g \in \bm{\mathcal{M}}_{\rm ele}}^{}{\left| \boldsymbol{\hat{H}}(g) - \boldsymbol{H}_{\rm gt}(g) \right|},
      \label{eq:elevation_loss}
\end{equation}

\noindent where $\boldsymbol{H}_{\rm gt}$ is the ground-truth (GT) elevation map, $\bm{\mathcal{M}}_{\rm ele}$ is the set of valid grids.

For the texture branch, we use both L1 loss and SSIM loss to supervise the synthesized image:
\begin{equation}
    \mathcal{L}_{\rm rgb} = \sum_{i \in \bm{\mathcal{I}}}\lambda | \boldsymbol{\hat{I}}_{i} - \boldsymbol{I}_{i}| +(1-\lambda)(1-{\rm SSIM}(\boldsymbol{\hat{I}}_{i},\boldsymbol{I}_{i})),
\end{equation}

\noindent where $\bm{\mathcal{I}}$ is the set of actual images under different poses. $\lambda$ is the weight. Since there may be empty areas in the synthesized image, the loss only considers the splattered pixels.

\begin{figure}[t]
  \centering
   \includegraphics[width=0.7\linewidth]{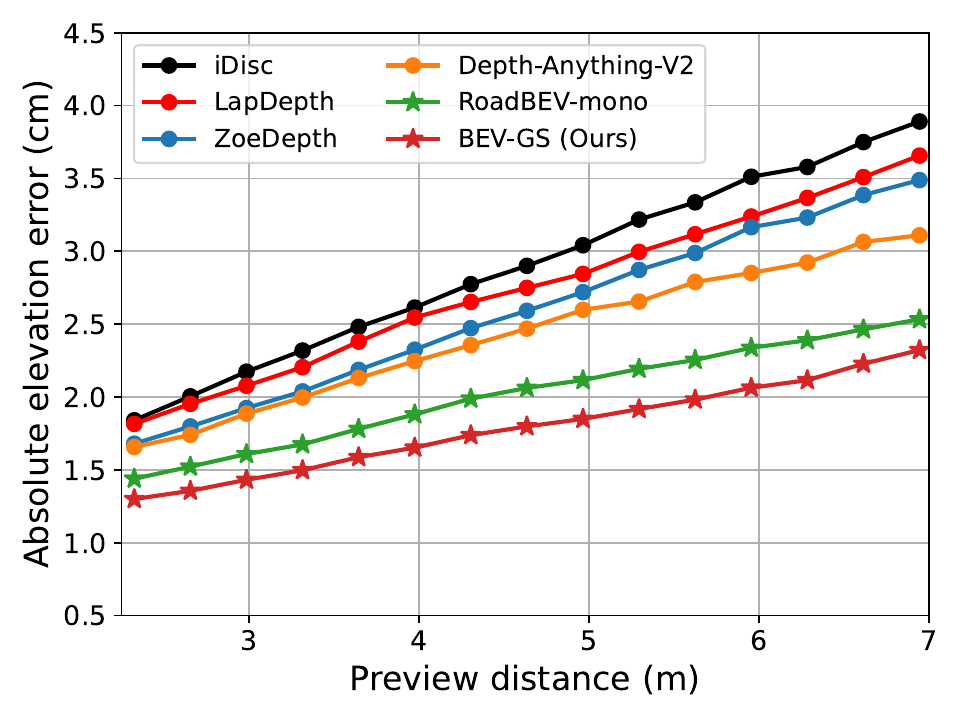}
   \vspace{-15pt}
   \caption{Comparison of segment-wise elevation errors with the SOTA depth estimation and BEV models.}
   \label{fig-ele2}
\end{figure}
\section{Experiments}

In this section, we demonstrate the excellent performance of our method through comprehensive experiments and highlight its superiority compared to both existing geometry and texture reconstruction methods.

\subsection{Experimental Settings}
\noindent \textbf{Dataset.} We conduct experiments on the RSRD dataset \cite{zhao2024road}, which is a real-world dataset focusing on road surface reconstruction. It offers high-accuracy data under various road conditions, including stereo images, point clouds, and processed GT elevation maps. The scenarios cover common road types, including a significant number of potholes and cracked road surfaces. We partitioned the RSRD dataset into a training set containing 926 samples and a test set containing 285 samples. The resolution of the input image is $H=528$ and $W=960$. 

\noindent \textbf{Evaluation Metrics.}
Following RoadBEV \cite{roadbev}, we calculate the average absolute error (AEE), root mean square error (RMSE), and the percentage of grids with absolute error exceeding 5mm ($>$ 5mm) between the predicted and GT elevation to evaluate geometry reconstruction performance. For texture metrics, due to the absence of GT colors of road surface in BEV, we evaluate texture reconstruction through perspective view rendering. We calculate the peak signal-to-noise ratio (PSNR), structural similarity index measure (SSIM), and the learned perceptual image patch similarity (LPIPS) between the synthesized and observed images.

\noindent \textbf{Implementation Details.}
\label{Implementation Details} 
The image feature maps have $C=128$ channels. The grid interval of road area is set as 3 cm, resulting in grid size $N^g_x=64$, and $N^g_y=164$. The elevation range is 20 cm above or below the zero reference plane, i.e., $h_{\rm min}$=-20 cm, and $h_{\rm max}$=20 cm. Each grid has $N^g_z=20$ elevation anchors for geometry feature query. There are $N_b=40$ bins for elevation offset classification. 

For higher resolution and more realistic rendering, the grid size of texture representation is four times that of the geometry, i.e., $N^t_x=256$, and $N^t_y=656$. The order of spherical harmonics is $L$=1. The scale of Gaussians $\boldsymbol{S}$ is initialized to [0.002, 0.002, 0.002]${}^{T}$, the opacity $\sigma$ is initialized to 1.0.
The weight $\lambda$ in $\mathcal{L}_{\rm rgb}$ is 0.5. In the training phase, we use stereo left and right images in RSRD for rendering supervision, i.e., $\bm{\mathcal{I}} = \{\boldsymbol{I}_l, \boldsymbol{I}_r\}$. The prediction network is trained for 50 epochs with AdamW optimizer. Batch size is 8. The maximum learning rate is $5\times {10}^{-4}$ with linear decay. 
All experiments are conducted on a server equipped with an NVIDIA A100 GPU.

\begin{figure}[t] 
\center
\includegraphics[width=0.7\linewidth]{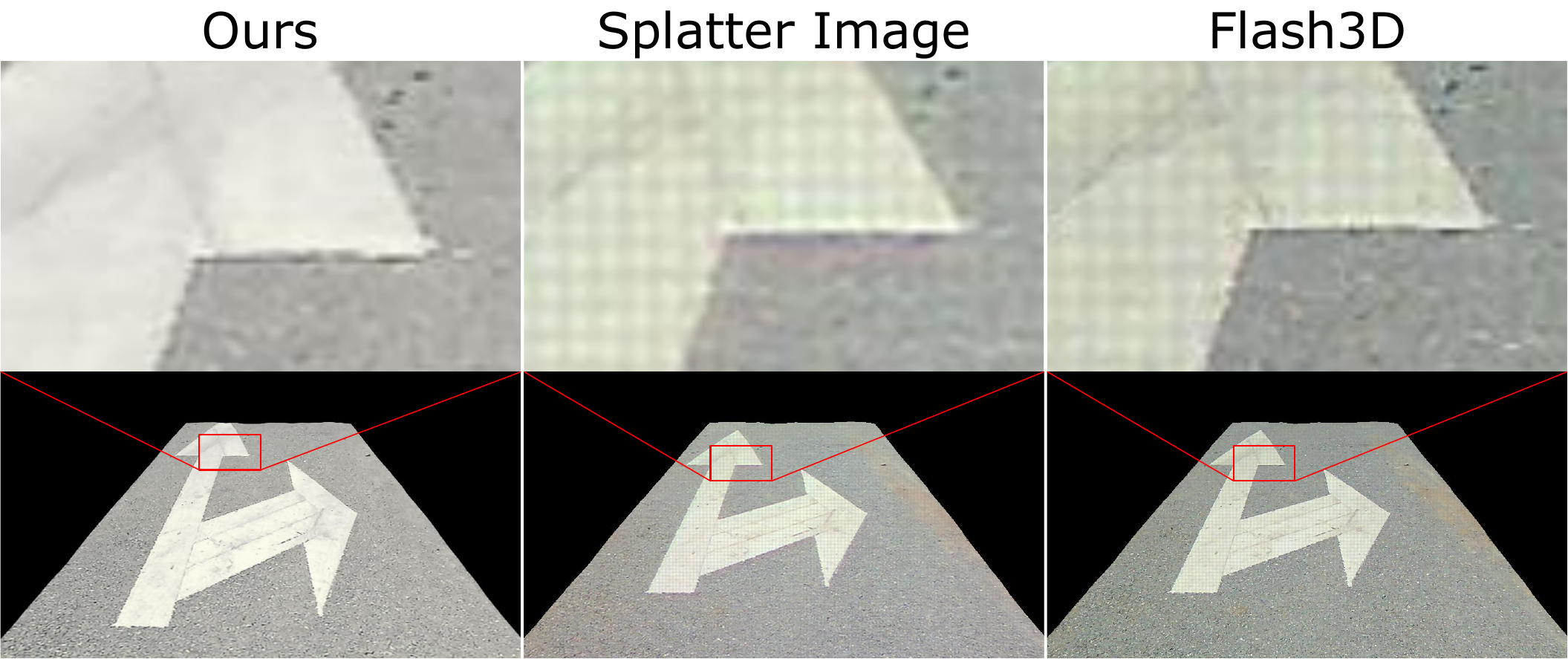}
\caption{Highlight of synthesized distant road surface.
}
\label{fig-opt}
\end{figure}

\begin{table}[t]
  \centering
    \begin{tabular}{c||c|c|c}
    \hline
    \makecell[c]{Methods} & \makecell[c]{PSNR $\uparrow$} & \makecell[c]{SSIM $\uparrow$} & \makecell[c]{LPIPS $\downarrow$} \\
    \hline
      Splatter Image\cite{szymanowicz2024splatter_image}   & 22.32 & 0.70 & 0.41  \\
      Flash3D \cite{szymanowicz2024flash3d}   & 23.27 & 0.71 & 0.41  \\
      \hline
      BEV-GS (Ours) & \textbf{28.36} & \textbf{0.77} & \textbf{0.14} \\
    \hline
    \end{tabular}%
    \caption{Comparison of novel view synthesis.  \textbf{Bold}: best.}
      \label{tab-rgb}%
\end{table}%

\noindent \textbf{Baselines.}
We compare our BEV-GS with state-of-the-art (SOTA)
single-image reconstruction methods. For scene geometry, we compare against method in BEV like RoadBEV \cite{roadbev}, and methods in perspective view like iDisc \cite{piccinelli2023idisc}, ZoeDepth \cite{ZoeDepth}, and Depth Anything V2 \cite{depth_anything_v2}. We select the DPT-BEiT-L as the backbone for ZoeDepth, and Swin-T for iDisc. The models above are all re-trained for 50 epochs with their default configurations on the RSRD dataset \cite{zhao2024road}. The maximum depth is set as 13m for the metric depth estimation models. For texture reconstruction, we compare against Splatter Image \cite{szymanowicz2024splatter_image} and Flash3D \cite{szymanowicz2024flash3d}.

\begin{table}[t]
  \centering
    \begin{tabular}{c|c||c|c|c}
    \hline
    \multicolumn{2}{c||}{Methods} & \makecell[c]{Infer.  FPS } & \makecell[c]{Rend.  FPS } & \makecell[c]{\# Params} \\
    \hline

    \multirow{8}{*}{\rotatebox{90}{Geometry only}} & iDisc \cite{piccinelli2023idisc} & 12.3 & - & 41M \\
    & LapDepth \cite{9316778} & 83.2 & - & 16M \\
    & PixelFormer \cite{Agarwal_2023_WACV} & 43.0 & - & 89M \\
    & AdaBins \cite{9578024} & 23.2 & - & 78M\\
    & ZoeDepth \cite{ZoeDepth} & 15.2 &-  & 345M\\
    & {Depth Anything V2\cite{depth_anything_v2}} & 6.6 &- & 335M \\
    & RoadBEV \cite{roadbev} & 26.8 & - & 27M \\
      \hline
    \multirow{3}{*}{\rotatebox{90}{Texture}} & Splatter Image\cite{szymanowicz2024splatter_image}   & 18.9 & 412 & 56 M  \\
    & Flash3D \cite{szymanowicz2024flash3d}   & 6.3 & 96 & 400 M  \\
    & BEV-GS (Ours) & 26.3 & 2061 & 8M \\
    \hline
    \end{tabular}%
\caption{Analysis of model runtime and complexity. `-' denotes no rendering.}
  \label{tab-time}%
\end{table}%

\subsection{Results and Comparison} \label{sota_comparison}
\noindent \textbf{Geometry Reconstruction Results.} Road elevation reconstruction results are shown in Table \ref{tab-ele}. For depth estimation methods in PV, we convert the recovered point clouds into BEV and generate elevation maps in the same style as building GT maps. Benefiting from direct elevation modeling, our BEV-GS achieves SOTA performance with a significant margin. The elevation AAE and RMSE reach 1.73cm and 1.94cm, reducing 10.4\% and 10.2\% than RoadBEV, respectively. Compared with monocular depth estimation, elevation in BEV is naturally more suitable for describing road geometry. Figure \ref{fig-ele} visualizes the predicted and GT elevation maps. 
BEV-GS provides more accurate and uniform road profiles than RoadBEV \cite{roadbev}, especially in areas with irregularities. The shape and edge of the potholes are all precisely captured. Recovered surface profiles on flat roads are also smooth without noise, avoiding wrong identifications or unnecessary actions in the subsequent tasks.

For more insights of reconstruction performance, we further visualize the segment-wise AAE along the longitudinal direction, as shown in Figure \ref{fig-ele2}. The entire grid is divided into 15 segments from near to far, with each segment representing a 33-centimeter-long region. By averaging the AAE values within each segment, a penetrative performance assessment is achieved. Our BEV-GS thoroughly outperforms the others in all segments, verifying its capability to capture fine-grained road structure. The maximum error can be limited within 2.5cm. It is also indicated that the elevation error increases with distance, which is still caused by the loss of texture details in farther areas.

\begin{figure}[t] 
\center
\includegraphics[width=0.7\linewidth]{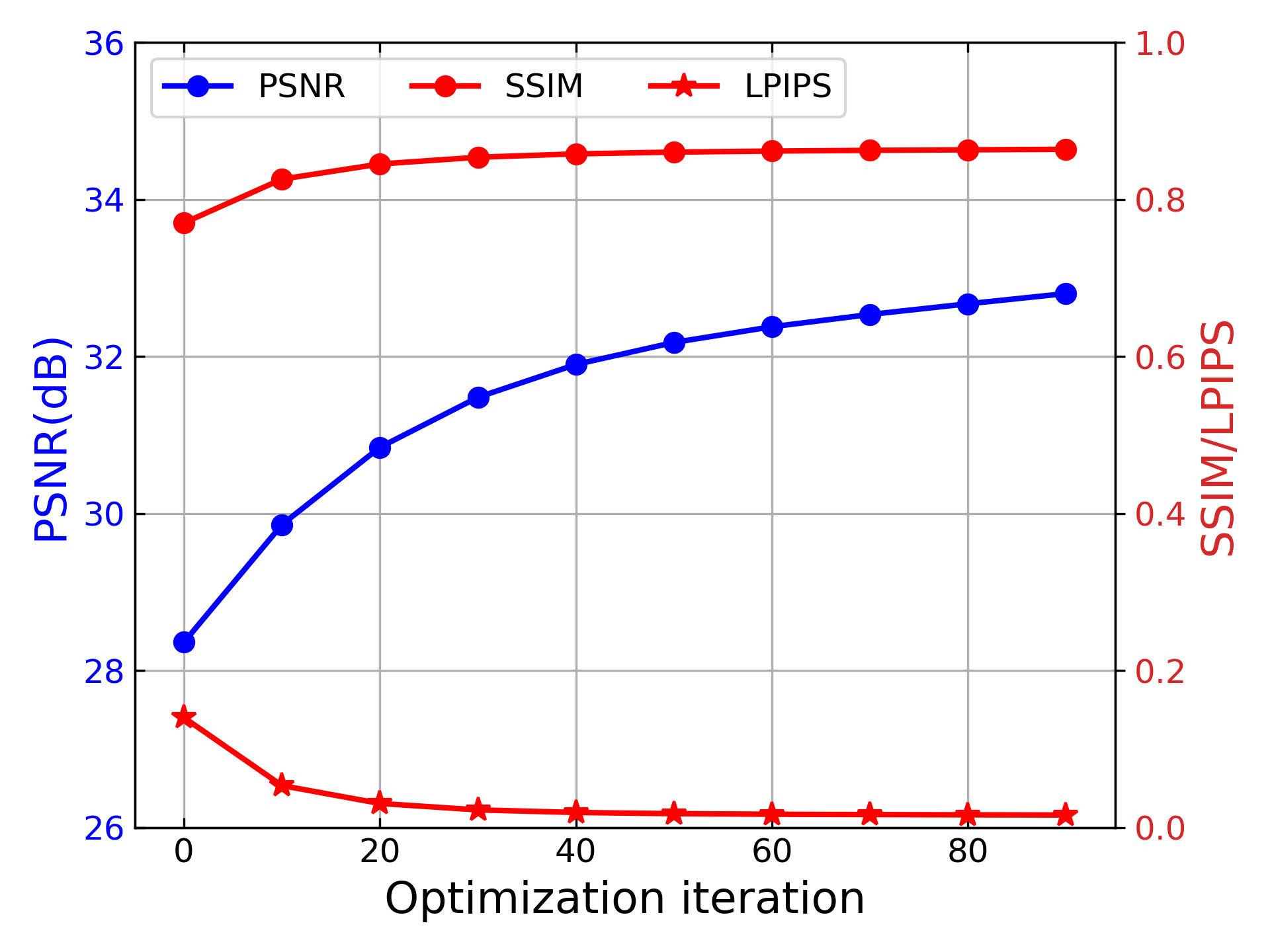}
\caption{Variation of synthesis metrics with the number of optimization iterations. Each iteration takes $11$ ms.
}
\label{fig-rgb2}
\end{figure}

\noindent \textbf{Texture Reconstruction Results.} 
Our BEV-GS also recovers detailed road color and texture, as shown in the last column of Figure \ref{fig-ele}. Fine structures of road cracks and potholes are all clearly retained without distortion, which are important for downstream road perception tasks such as condition understanding.
Table \ref{tab-rgb} presents the quantitative performance comparison of novel view synthesis. Our BEV-GS outperforms the very recent SOTA models significantly across all three metrics. Figure \ref{fig-rgb} visualizes the comparison of actual and synthesized novel views in typical road scenarios. 
The accurate geometry and BEV-based grid Gaussian representation contribute to globally consistent and high-quality novel view synthesis. Compared with the per-pixel Gaussian prediction methods, our method synthesizes images with more accurate correspondence, more realistic colors, and finer details. Figure \ref{fig-rgb2} highlights the rendering of distant road surfaces. Our method retains fine details, whereas others produce poor rendering images due to the sparsity of distant road Gaussians.

\noindent \textbf{Runtime Analysis.} 
Table \ref{tab-time} compares inference or rendering speed, as well as model complexity. Among the methods for road geometry recovery, BEV-GS achieves outstanding inference speed with much less parameters. For joint texture and geometry reconstruction, BEV-GS outperforms the others among all metrics. Benefit from BEV grid representation, the feed-forward prediction speed at 26 FPS implies real-time capability for practical applications.

\noindent \textbf{Test Time Optimization.} 
In addition to prediction, the scheme enables backward and iterative optimization during testing. The predicted geometry and texture properties are treated as initial parameters of GS optimizer, which further updates color, rotation, and opacity of Gaussians. The learning rates for zero and first order spherical harmonics, rotation, and opacity are 0.05, 0.001, 0.001, and 0.001, respectively. We only use the current frame for supervision with the color loss $\mathcal{L}_{\rm rgb}$. Figure \ref{fig-opt} illustrates the performance enhancement with further optimization. The synthesis metrics are all improved, converging around 60 iterations. Test-time optimization significantly enhances texture reconstruction quality, albeit at the cost of increased processing time. In applications, the decision to employ test-time optimization and the number of iterations can be based on the trade-off between reconstruction quality and real-time requirements.

\begin{table}[t]
  \centering
    \begin{tabular}{l||cccccc}
    \hline
    \makecell[c]{Configurations} &\makecell[c]{a. w/ shared backbone }&\makecell[c]{b. w/o elevation guidance } &\makecell[c]{c. full method} \\
    \hline
    AAE (cm) $\downarrow$  & 1.89 & \textbf{1.73} & \textbf{1.73} \\
     RMSE (cm) $\downarrow$  & 2.13 &\textbf{ 1.94} & \textbf{1.94} \\
    $>$5mm (\%) $\downarrow$& 82.1 & \textbf{80.2} & \textbf{80.2} \\
   PSNR (dB) $\uparrow$ & 24.90 & 25.00 & \textbf{28.36} \\
    SSIM$\uparrow$ &0.48 & 0.48 & \textbf{0.77} \\
    LPIPS $\downarrow$& 0.28 & 0.26 & \textbf{0.14} \\
    \hline
    \end{tabular}%
\caption{Ablation Study. Results for ablating different design configurations of our method.}
  \label{tab-abl}%
\end{table}%

\subsection{Ablation Studies} \label{Ablation studies}
\noindent \textbf{Effectiveness of Geometry-Texture Separation.} Table \ref{tab-abl} a. presents the results of utilizing a shared feature extractor for geometry and texture. Both geometry and texture reconstruction metrics exhibit a noticeable decline. The geometry and texture exhibit distinct feature patterns. When a shared feature extractor is employed, geometry and texture features become coupled, leading to mutual interference. 
The proposed two-branch design decouples these aspects, facilitating improved learning of geometry and texture.

\noindent \textbf{Effectiveness of Elevation Guidance.} Table \ref{tab-abl} b. demonstrates the results of BEV texture feature queries without elevation guidance.
With hierarchical BEV feature queries, texture features from different elevations tend to overlap. In contrast, elevation-guided BEV feature queries not only retrieve more accurate texture features but also significantly reduce the computational burden, requiring only one query per grid feature. The experimental results underscore the enhancement in texture reconstruction results facilitated by elevation guidance.
\section{Conclusion}

We propose BEV-GS, a novel real-time single-frame road surface reconstruction method based on feed-forward Gaussian splatting. 
By introducing a feed-forward geometric and texture prediction network based on BEV representation, our method directly predicts road elevation and Gaussian parameters from a single image. Extensive experiments across various complex road surfaces demonstrate that our method efficiently and accurately reconstructs road geometry and texture from a single image, while also providing high-quality novel-view synthesis. It has great potential for online applications like road condition preview and offline tasks such as autonomous driving testing. 


\bibliographystyle{unsrt}  
\bibliography{references} 

\appendix

\clearpage
\setcounter{page}{1}

\section{Overview}
\label{sec: Overview}
In this document, we present more details and several extra results. In Section~\ref{sec: implementation}, we elaborate on the implementation details of our method. In Section~\ref{sec: result}, we presented the road surface reconstruction results for different grid resolutions and test time optimization.

\section{Further Implementation Details}
\label{sec: implementation}

\subsection{Hyperparameters}
The resolution of the input image is $H=528$ and $W=960$. The image feature maps have $C=128$ channels and the resolution is $132 \times 240$. We predefine a horizontal region of interest in front of the vehicle with size $1.9{\rm m} \times 5.0{\rm m}$. The grid interval of road area is set as 3 cm, resulting in grid size $N^g_x=64$, and $N^g_y=164$. For higher resolution and more realistic rendering, the grid size of texture representation is four times that of the geometry, i.e., $N^t_x=256$, and $N^t_y=656$. The size of the Gaussian grid is then upsampled as $512\times 1312$. The initial learning rate is $5\times {10}^{-4}$ and linearly decays to one-twentieth of the initial value.

\subsection{Evaluation Metrics}
Following RoadBEV \cite{roadbev}, we calculate the average absolute error (AEE), root mean square error (RMSE), and the percentage of grids with absolute error exceeding 5mm ($>$ 5mm) between the predicted and GT elevation to evaluate geometry reconstruction performance. Their formulas are as follows:
\begin{equation}
    e(g) = \boldsymbol{\hat{H}}(g) - \boldsymbol{H}_{\rm gt}(g),
\end{equation}
\begin{equation}
    \text{AEE} = \frac{1}{\left|\bm{\mathcal{M}}_{\rm ele} \right|}\sum_{g \in \bm{\mathcal{M}}_{\rm ele}}^{}{\left| e(g) \right|},
\end{equation}
\begin{equation}
    \text{RMSE} = \sqrt{ \frac{1}{\left|\bm{\mathcal{M}}_{\rm ele} \right|}\sum_{g \in \bm{\mathcal{M}}_{\rm ele}}^{}{ e(g)^2}},
\end{equation}
\begin{equation}
    > \text{5mm} = \frac{1}{\left|\bm{\mathcal{M}}_{\rm ele} \right|}\sum_{g \in \bm{\mathcal{M}}_{\rm ele}}^{}{u(\left| e(g) \right|- 0.005)} \times 100 \%,
\end{equation}
where $\boldsymbol{\hat{H}}_{\rm gt}$ and $\boldsymbol{H}_{\rm gt}$ are the predicted and ground-truth (GT) elevation map, $\bm{\mathcal{M}}_{\rm ele}$ is the set of valid grids. $\left|\bm{\mathcal{M}}_{\rm ele} \right|$ represents the number of valid grids. $u(\cdot)$ stands for unit step function.

For texture metrics, due to the absence of GT colors of road surface in BEV, we evaluate texture reconstruction through perspective view rendering. We calculate the peak signal-to-noise ratio (PSNR), structural similarity index measure (SSIM), and the learned perceptual image patch similarity (LPIPS) between the synthesized and observed images.
\section{Further Ablation Analysis and Visualization}
\label{sec: result}

\subsection{Test Time Optimization.}
\begin{figure*}[t] 
\center
\includegraphics[width=1.0\textwidth]{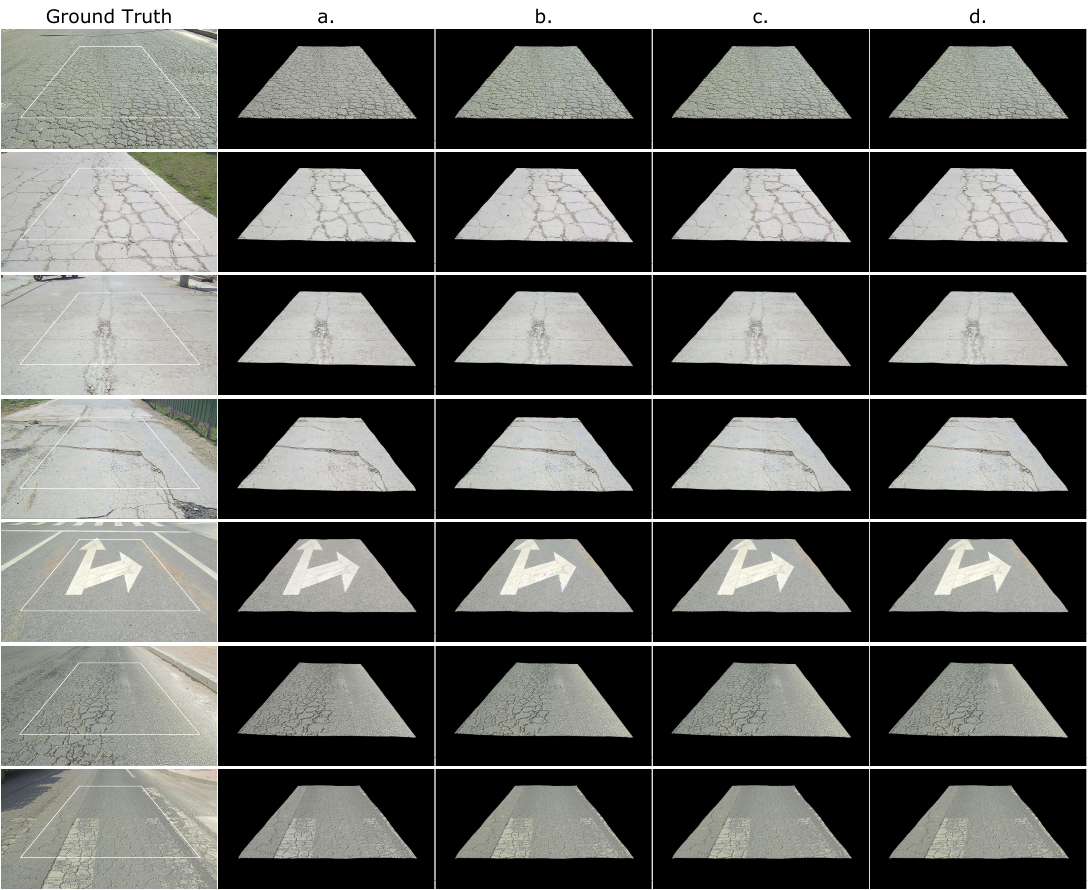}
\caption{Visualization of road synthesis with different optimization iterations.
}
\label{fig-2}
\end{figure*}

In addition to prediction, the scheme enables backward and iterative optimization during testing. The experimental section of the main text has already shown the variation of reconstruction metrics with the number of iterations. Here, we provide additional detailed quantitative results and visualizations, as presented in Table \ref{tab-2} and Figure \ref{fig-2}
Test-time optimization significantly enhances texture reconstruction quality, albeit at the cost of increased processing time. In applications, the decision to employ test-time optimization and the number of iterations can be based on the trade-off between reconstruction quality and real-time requirements.

\begin{table}[h]
  \centering
    \begin{tabular}{c|c||c|c|c}
    \hline
    & Opt. Iter. & \makecell[c]{PSNR $\uparrow$} & \makecell[c]{SSIM $\uparrow$} & \makecell[c]{LPIPS $\downarrow$} \\
    \hline
     a.& 0 & 28.361 & 0.772 & 0.136\\
     b.& 30 & 30.843 & 0.854 & 0.022\\
     c.& 60 & 32.178 & 0.862 & 0.017\\
     d.& 90 & \textbf{32.800} & \textbf{0.864} & \textbf{0.016}\\
    \hline
    \end{tabular}%
    \caption{Variation 
 of synthesis metrics with the number of optimization iterations. Each iteration takes 11 ms.  \textbf{Bold}: best.}
      \label{tab-2}%
\end{table}%

\subsection{Grid Resolution Ablation}
\begin{figure*}[t] 
\center
\includegraphics[width=1.0\textwidth]{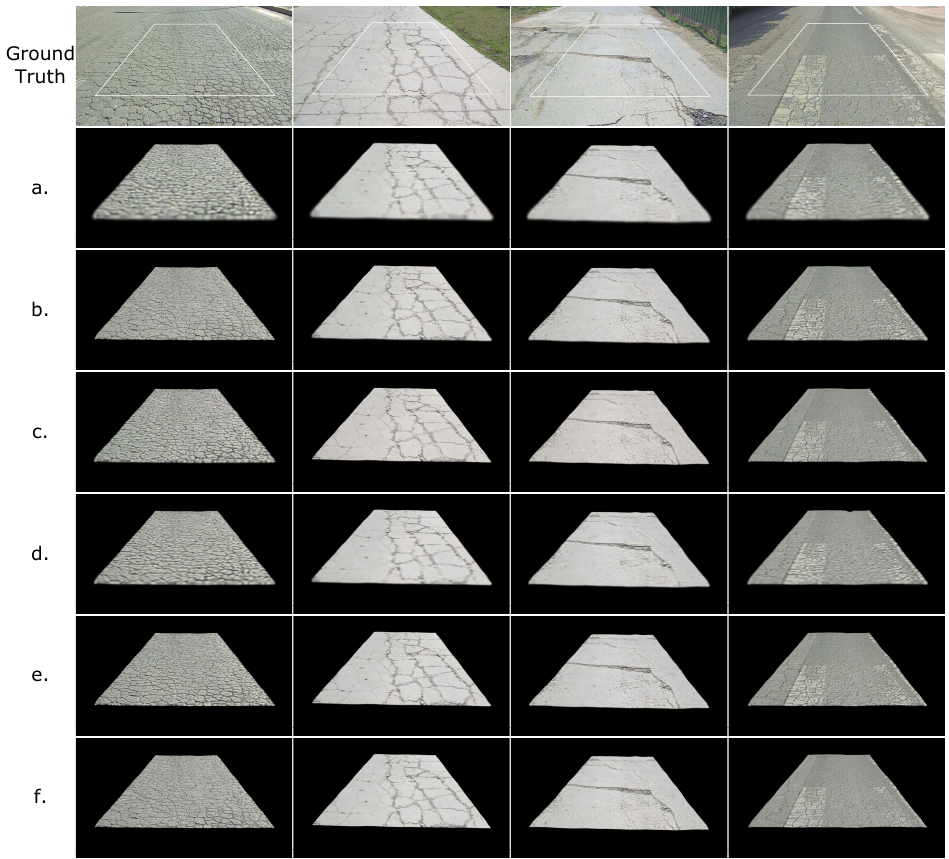}
\caption{Visualization of road synthesis with different texture grid resolution and Gaussian grid resolution.
}
\label{fig-1}
\end{figure*}

The resolution of texture grids and Gaussian grids will impact the effect of texture reconstruction. Experiments with different grid resolutions were conducted to investigate this, and the results are presented in Table \ref{tab-1}. It was observed that higher resolutions of both texture grids and Gaussian grids lead to better reconstruction effects. Figure \ref{fig-1} illustrates the visual reconstruction of road surface colors at various grid resolutions. It is evident that higher-resolution grids can capture more detailed road surface textures. Furthermore, when the resolution of the texture grids is consistent, initializing higher-resolution Gaussian grids through upsampling can also enhance the reconstruction performance.

\begin{table*}[t]
  \centering
    \begin{tabular}{c|c||c|c|c|c|c}
    \hline
    \multicolumn{2}{c||}{Configurations} & Texture Grid & Gaussian Grid &\makecell[c]{PSNR $\uparrow$} & \makecell[c]{SSIM $\uparrow$} & \makecell[c]{LPIPS $\downarrow$} \\
    \hline
           \multirow{3}{*}{\rotatebox{90}{w/o  up.}} &a.& $64 \times 164$& $64 \times 164$ & 25.46 & 0.50 & 0.42  \\
                 & b. & $128 \times 328$&$128 \times 328$ & 26.14 & 0.59 & 0.32  \\
       & c. &$256 \times 656$ & $256 \times 656$& 27.63 & 0.73 & 0.18  \\
      \hline
   \multirow{3}{*}{\rotatebox{90}{w/  up.}}& d.  &$64 \times 164$ &$128 \times 328$ & 25.58 & 0.52 & 0.36 \\
  & e. &$128 \times 328$ &$256 \times 656$ & 26.39 & 0.63 & 0.26 \\
& f. &$256 \times 656$  & $512 \times 1312$&\textbf{28.36} & \textbf{0.77} & \textbf{0.14} \\
    \hline
    \end{tabular}%
    \caption{Results for ablating different texture grid resolution and Gaussian grid resolution. `up.' stands for upsampling. \textbf{Bold}: best.}
      \label{tab-1}%
\end{table*}%

\end{document}